\RequirePackage{ifthen}
\newcommand{\ite}{\ifthenelse}
\newcommand{\bl}{\boolean}
\newboolean{prep}
\setboolean{prep}{true} 
\ite{\bl{prep}}{
\documentclass[preprint,preprintnumbers,amsmath,amssymb,a4paper]{revtex4}
}
{
\documentclass[twocolumn,preprintnumbers,a4paper]{revtex4}
}

\usepackage{amsmath}
\usepackage{amssymb}
\usepackage{subfigure}
\usepackage{epsfig}

\begin{document}

\preprint{APS/123-QED}

\title{High Curie temperatures in (Ga,Mn)N from Mn clustering}

\author{Teemu Hynninen}
\affiliation{%
COMP/Laboratory of Physics, 
Helsinki University of Technology, P.O. Box 1100, 02015 HUT, Finland.
}
\author{Hannes Raebiger}%
 \email[Corresponding author, e-mail: ]{hra@fyslab.hut.fi}
\affiliation{%
COMP/Laboratory of Physics, 
Helsinki University of Technology, P.O. Box 1100, 02015 HUT, Finland.
}%
\author{Andr\'es Ayuela}%
\affiliation{%
Donostia International Physics Centre (DIPC), 
P.O. Box 1072, 20018 Donostia - San Sebastian, Spain
}%
\author{J. von Boehm}%
\affiliation{%
COMP/Laboratory of Physics, 
Helsinki University of Technology, P.O. Box 1100, 02015 HUT, Finland.
}%

\date{\today}

\begin{abstract}
   The effect of microscopic Mn cluster distribution on the Curie
temperature ($T_C$) of (Ga,Mn)N is studied using 
density-functional calculations together with the mean field approximation.
We find that the calculated $T_C$ depends crucially
on the microscopic cluster distribution,
which can explain the abnormally large variations in 
experimental $T_C$ values from a few K to well above room temperature. 
The partially dimerized Mn$_2$--Mn$_1$
distribution is found to give the highest $T_C > 500$ K, and in
general, the presence of the Mn$_2$ dimer has a tendency to enhance $T_C$.
The lowest $T_C$ values close to zero are obtained for the 
Mn$_4$--Mn$_1$and Mn$_4$--Mn$_3$
distributions.
\end{abstract}

\maketitle

   The discovery of ferromagnetism in (In,Mn)As~\cite{munekata-ea-1989}
sprouted a family of III-V based diluted magnetic semiconductors.
According to the mean-field Zener theory (Ga,Mn)N should have
the highest Curie temperature $T_C$ ($\simeq 400$ K) among
these prospective III-V spintronics materials~\cite{dietl-ea-2000}. However, the
experimental $T_C$ values for (Ga,Mn)N range from 10 to
940 K~\cite{overberg-ea-2001,sasaki-ea-2002,thaler-ea-2002,dhar-ea-2003a}.
First principles calculations combined with the mean field approximation
have given rather high $T_C$ values of about 
270-350 K~\cite{sandratskii-ea-2005,sato-ea-2003a,sandratskii-ea-2004},
while more sophisticated statistical approaches including
percolation effects and/or magnetic fluctuations give
significantly lower Curie 
temperatures~\cite{xu-ea-2005,bergkvist-ea-2004,sato-ea-2004,bouzerar-ea-2005}.
The
substantial variation of the experimental $T_C$ values is suggested
to be due to clustering of Mn atoms~\cite{dhar-ea-2003a,bouzerar-ea-2004}
and the formation of giant magnetic moments at the Mn
clusters~\cite{rao-jena-2002}. Consistently with this, it is found that it is
energetically favourable for substitutional Mn atoms to cluster
around N atoms~\cite{schilfgaarde-mryasov-2001,rao-jena-2002}. Also,
Mn clusters are observed directly at high Mn
concentrations~\cite{shon-ea-2003,giraud-ea-2004,martinez-criado-ea-2004}.
Although in many theoretical calculations
Mn clustering is generally found to decrease 
$T_C$'s in III-V 
materials~\cite{sandratskii-bruno-2004,raebiger-ea-2004,xu-ea-2005,raebiger-ea-2005,sandratskii-ea-2005}, 
we show in this Letter
by using first principles calculations and the mean field theory
that Mn clustering may also act beneficially by increasing $T_C$.
More generally, we find that the stability of the ferromagnetic order
depends crucially on the microscopic cluster distribution: small Mn$_2$
clusters may increase $T_C$ whereas Mn$_4$ clusters may
reduce $T_C$ down to a few Kelvins (by an Mn$_i$ cluster with $i = 2,3,4$
we mean $i$ Mn atoms surrounding the same central neigbouring N atom).

    The presence of Mn clusters in (Ga,Mn)N is unavoidable at the Mn
concentrations that give high $T_C$ values. This is obvious from
Table~\ref{cl} where Mn cluster distributions are given when the Ga atoms
are substituted randomly by the Mn atoms at various relevant
Mn concentrations. Already at $x = 0.02$ (counting single Mn atoms as
Mn$_1$ clusters) the Mn$_2$ cluster portion exceeds
10 \% and grows fast up to 40 \% with increasing $x$. Also the 
Mn$_3$ portion increases markedly with increasing $x$ whereas the
Mn$_4$ cluster portion remains less than 0.5 \% even for
highest Mn concentration of $x = 0.14$. 
\newcommand{\captabi}{Mn cluster distribution as a function of $x$ when
the Ga atoms are substituted randomly by the Mn atoms.
The portions are given in units of \%
(counting single Mn atoms as Mn$_1$ clusters).
}
\newcommand{\tabi}[1]{
\begin{table}
\caption{#1}
\label{cl}
\begin{tabular}{l c c c c}
\hline\hline
 $x$ & Mn$_1$ & Mn$_2$ & Mn$_3$ & Mn$_4$ \\
\hline
 0.02 & 89 & 11 & 0 & 0 \\
 0.06 & 71 & 27 & 2 & 0 \\
 0.10 & 58 & 37 & 5 & 0 \\
 0.14 & 49 & 42 & 9 & 0 \\
\hline\hline
\end{tabular}
\end{table}
}
\ite{\bl{prep}}{}{\tabi{\captabi}}
To investigate the microscopic cluster distribution effects on the Curie temperature
we perform spin-polarized total energy
density-functional calculations using the projector augmented wave method as
implemented in the VASP code~\cite{kresse-furthmuller-1996}.
Our supercells contain two Mn$_i$ clusters ($i=1,2,3,4$) at maximum separation,
thus representing a uniform distribution of
clusters.
The exchange-correlation is
approximated as follows. In the first approach we use the generalized gradient
approximation (GGA-PW91). In the second approach we use the local spin density
approximation plus $U$ (L + $U$) with $U$ chosen as 3 eV and used only for Mn.
The second approach tries to compensate the self-interaction which is
present in the local density approximation and which is known to be
substantial in (Ga,Mn)N~\cite{filippetti-ea-2004}. 
The plane-waves cut-off
is 425 eV. We use supercells
of 48, 72, or 108 atoms ($3\times2\times2$, $3\times3\times2$, or
$3\times3\times3$ primitive
wurzite unit cells, respectively) and
the $4\times6\times4$, $4\times4\times4$ or $4\times4\times3$ $\vec k$-meshes
for the Brillouin zone sampling, respectively.
We use throughout the same lattice parameters $a = 3.217$ \AA{} and
$c:a = 1.631$ obtained from geometry optimization of the wurzite GaN
crystal. 
The atomic positions are fully relaxed in the 48 and 72 atom supercells,
and in the 108 atom supercells the atomic positions are fixed to the respective ones
of the smaller relaxed supercells.
The $T_C$ values are roughly estimated from the mean-field
expression~\cite{sato-ea-2003a,kurz-ea-2002,turek-ea-2003} 
\begin{equation}
T_C = \frac{2}{3 k_B}\frac{\Delta}{N},
\label{eq:tc}
\end{equation}
where $N$ is the number of Mn clusters in the supercell fixed to 2,
and $\Delta$ is the difference in supercell total energy between the anti-parallel
and parallel alignments.

First we find,
in agreement with
Ref.~\cite{schilfgaarde-mryasov-2001},
that the Mn$_i$ cluster ($i$ = 2,3,4) is energetically the most stable
configuration of $i$ Mn atoms. 
This suggests that the
Mn$_2$, Mn$_3$ and Mn$_4$ cluster portions may in reality be
even larger 
than the ones based on random substitution given in Table~\ref{cl}.
We also find in agreement with
Refs~\cite{rao-jena-2002,schilfgaarde-mryasov-2001} that large stable
magnetic moments are formed at the clusters, and therefore intercluster
interactions determine the ferromagnetic order.

\newcommand{\captabii}{%
Spin-flip energies ($\Delta$) and Curie temperatures ($T_C$) for
two clusters in the supercell. 
The fourth and fifth columns are GGA
calculations, the sixth and seventh columns L+$U$ calculations (denoted by the
index ``+$U$''). $d$ is the
minimum distance between the Mn atoms belonging to the different centers.
}
\newcommand{\tabii}[1]{
\begin{table}
\caption{#1}
\label{sfe}
\begin{tabular}{l c c c c c c c}
\hline\hline
System & & $x$ (\%) & $d$ (\AA) & $\Delta$ (meV) & $T_C$ (K) & $\Delta^{+U}$ & $T_C^{+U}$ \\
\hline
Mn$_1$ - Mn$_1$ $^a$  & & 5.6 & 7.65 & 10 & 39 & 1 & 4 \\
Mn$_1$ - Mn$_1$ $^b$  & & 8.3 & 6.15 & 92 & 355 & - & - \\
Mn$_2$ - Mn$_1$ $^a$  & & 8.3 & 6.13 & 133 & 514 & 160 & 618 \\
Mn$_3$ - Mn$_1$ $^a$  & & 11.1 & 6.13 & 76 & 294 & 117 & 453 \\
Mn$_3$ - Mn$_1$ $^c$  & & 7.4 & 9.28 & 33 & 128 & - & -\\
Mn$_4$ - Mn$_1$ $^c$  & & 9.3 & 7.66 & 7 & 27 & 10 & 39\\
Mn$_4$ - Mn$_3$ $^c$  & & 13.0 & 6.19 & 1 & 4 & - & -\\
\hline\hline
\multicolumn{8}{l}{$^a$ 72 atoms/supercell (SC), $^b$ 48 atoms/ SC, $^c$ 108 atoms/SC} 
\end{tabular}
\end{table}
}
\ite{\bl{prep}}{}{\tabii{\captabii}}

\newcommand{\capdosi}{
Densities of states 
together with Mn $d$ and N $p$ orbital projections
for most likely cluster arrangements in (Ga,Mn)N. 
(a) Uniform Mn
distribution for $x = 0.056$, in the inset for $x = 0.083$.
(b) Partially dimerized Mn$_2$ - Mn$_1$ for $x = 0.083$, in the inset
a uniform Mn$_2$ distribution for $x = 0.056$.
}
\newcommand{\dosi}[2]{
\begin{figure}[htb!]
\begin{center}
\epsfig{file=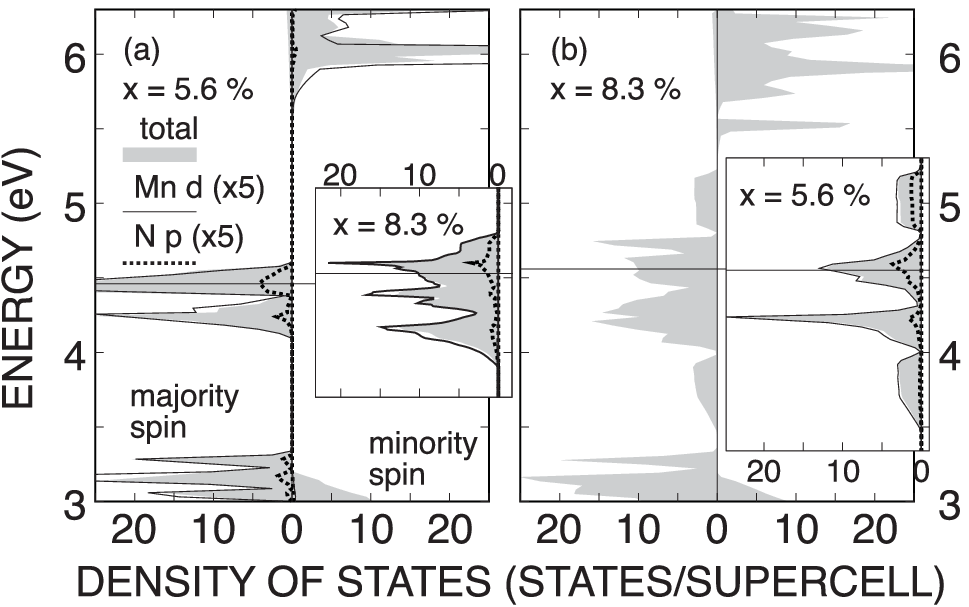,width=#1}
\caption{#2}
\label{dos1}
\end{center}
\end{figure}
}
\ite{\bl{prep}}{}{\dosi{\linewidth}{\capdosi}}

\newcommand{\capdosii}{Densities of states for (Ga,Mn)N. (a) Mn$_4$ - Mn$_1$
for $x = 0.111$. (b) Uniform Mn$_4$ distribution for $x = 0.093$. For further
information see the caption of Fig.~\ref{dos1}.
}
\newcommand{\dosii}[2]{
\begin{figure}[htb!]
\begin{center}
\epsfig{file=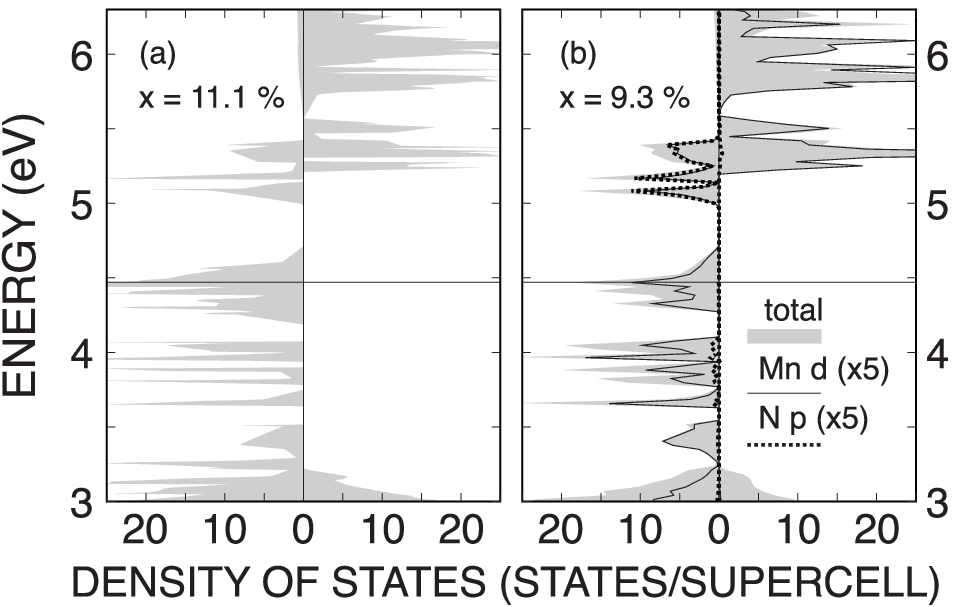,width=#1}
\caption{#2}
\label{dos2}
\end{center}
\end{figure}
}
\ite{\bl{prep}}{}{\dosii{\linewidth}{\capdosii}}

We now address the clustering effects on the Curie temperature
within the GGA and the mean field approximation.
In the case of
the (nearly) uniform distribution of the single Mn atoms, 
when the Mn
concentration $x$ increases from 0.056 to 0.083 (and the shortest Mn - Mn
distance decreases from 7.65 to 6.15 \AA) 
the calculated
$T_C$ increases sharply from 39 to 355 K,
as seen in the first two rows of~Table~\ref{sfe},
qualitatively in agreement with Ref.~\cite{sandratskii-ea-2005}. 
The increased magnetic Mn - Mn coupling is reflected in the density
of states (DOS) (given in Fig.~\ref{dos1} (a))
where the width of the midgap Mn $d$ type majority spin DOS
increases from 0.5 to 1 eV (see the inset in Fig.~\ref{dos1} (a)).
    Next we allow partial
dimerization at the same concentration of $x =0.083$. Two thirds of the
Mn atoms are allowed to form Mn$_2$ clusters in the 
plane perpendicular to the $\bf c$ axis.
This causes a further
\textit{increase} of $T_C$ up to 514 K given in Table~\ref{sfe}.
In Fig.~\ref{dos1} (b) for $x =0.083$ the partial dimerization is seen to cause
further broadening of the midgap majority spin DOS into a 1.7 eV wide joined
structure with an increased $p$ - $d$ hybridization. This atomic configuration
offers the highest $T_C$ we have found. 
In more detail, 
the DOS of a uniform Mn$_2$ distribution for $x =0.056$ is given 
in the inset in Fig.~\ref{dos1} (b), and comparison with this inset shows that the
structure in the DOS is mainly determined from the Mn$_2$ DOS.

    To see whether clustering from Mn$_2$ to Mn$_3$
could raise the Curie temperature further, above 514 K, we add a third atom to the Mn$_2$
cluster in the plane perpendicular to the $\bf c$ axis. 
However, in this case 
the clustering decreases $T_C$ to 294 K in agreement with
earlier experience~\cite{sandratskii-ea-2005,xu-ea-2005,sandratskii-bruno-2004,raebiger-ea-2004,raebiger-ea-2005}. 
The midgap DOS splits into four
distinct subbands (with three gaps) still with a clear $p$ - $d$ hybridization.
When the distance between the Mn$_3$
cluster and the single Mn atom is increased 
(the distance between the central N atom of the cluster and the single Mn atom grows from 5.90
to 9.08 \AA) $T_C$ decreases further to 128 K (see Table~\ref{sfe}), as
expected.

Finally, we complete the Mn$_3$ cluster to a symmetric Mn$_4$
tetramer cluster and find that $T_C$ diminishes down to 27 K (Table~\ref{sfe}).
The corresponding DOS shown in Fig.~\ref{dos2} (a) consists of the
narrow peak of about 0.5 eV at the Fermi level and of several other
separated peaks. The total width of the original majority spin midgap DOS
of 0.5 eV has grown to 2.1 eV which exceeds the calculated gap. The upper part overlaps
with the minority spin conduction band edge. The reason for the low $T_C$
may be understood by comparing with the DOS of a uniform Mn$_4$ distribution
($x = 0.093$) shown in Fig.~\ref{dos2} (b). It is immediately obvious that
the essential $p$ - $d$ hybridization is completely missing
from the narrow peak at the Fermi level. The central narrow subband
is of purely Mn $d$ type because the N $p$ orbitals are mainly used
for the Mn$_4$ intracluster bonding which makes the magnetic coupling of
the Mn$_4$ clusters to other Mn clusters (in this case to Mn$_1$) weak.
Summarizing, we find that the presence of the Mn$_4$
clusters leads to a low Curie temperature. 
This explains also why in the Mn$_4$ - Mn$_3$ case $T_C$ drops to
4 K although the Mn concentration is as high as $x = 0.13$
(Table~\ref{sfe}). On the other hand, the presence of the Mn$_2$ clusters
has a tendency to increase $T_C$.

    The L + $U$ results differ from the GGA ones mainly in shifting the
midgap Mn $d$ type DOS towards the valence band maximum by 0.5 eV.
In the partial dimerization the Mn $d$ type DOS
merges with the top of the valence band causing
a growth of $T_C$ up to 618 K (Table~\ref{sfe}). 
This
behaviour begins to resemble that in (Ga,Mn)As and might be assigned to
the $p$ - $d$ type hole mediated ferromagnetic coupling.
In the Mn$_3$-Mn$_1$,  and
especially in the Mn$_4$-Mn$_1$ case, the
$p$ - $d$ type empty peak splits off decreasing $T_C$ 
to 453 and 39 K, respectively (Table~\ref{sfe}),
similarly as for (Ga,Mn)As in 
Ref.~\cite{raebiger-ea-2005}.
All in all it seems that the previous conclusions on clustering effects on the
Curie temperature hold also in the case when further corrections to
the electronic correlation beyond L + $U$ are included.

The mean field approximation used in this work is known to
overestimate Curie 
temperatures~\cite{xu-ea-2005,bergkvist-ea-2004,sato-ea-2004,bouzerar-ea-2005},
i.e. the dependence of $T_C$ on the total energy difference $\Delta$ 
may not be linear as in Eq.~(\ref{eq:tc}).
Nevertheless, high $\Delta$ will imply high $T_C$, and thus the general trends
presented for different microscopic cluster configurations should hold.
It is interesting to remark that a recent study considering dimerization effects
beyond the mean field approximation~\cite{bouzerar-ea-2004} 
leads to a similar conclusion 
as the present work.

      In conclusion, we find that the calculated Curie temperature depends crucially
on the microscopic cluster distribution. The partially dimerized
Mn$_2$-Mn$_1$ distribution is found to give the highest $T_C$. 
This suggests that the presence of the Mn$_2$ dimer provides a 
mechanism to enhance $T_C$.
On the other hand,
low $T_C$ values are obtained for the Mn$_4$-Mn$_1$ and Mn$_4$-Mn$_3$
distributions.
Thus we have shown that different cluster distributions may explain
the abnormally large variations in experimental Curie temperatures.

   Acknowledgements. This work has been supported by the Academy of Finland
through the Center of Excellence Program (2000-2005). A.A. acknowledges the
financial support from the Basque Government by the ETOREK program called
NANOMAT. The authors thank Acad. Prof. R. M. Nieminen, Prof. K. Saarinen,
Prof. M. J. Puska, Dr. M. Alava, Mr F. Tuomisto, and Mr R. Oja for many valuable
discussions. The computing resources from the Center for Scientific
Computing (CSC) are acknowledged.

\ite{\bl{prep}}{\pagebreak}{}

\ite{\bl{prep}}{
\pagebreak
TABLE~\ref{cl}
\captabii

\bigskip

TABLE~\ref{sfe}
\captabii

\bigskip

FIG~\ref{dos1}
\capdosi

\bigskip

FIG~\ref{dos2}
\capdosii

\clearpage

\tabi{\captabi}

\clearpage

\tabii{\captabii}

\clearpage

\dosi{8cm}{}

\clearpage

\dosii{8cm}{}
}{}
\end{document}